\documentclass{ws-ijmpb}

\usepackage{graphicx}

\usepackage{cite}

\begin{document}

\markboth{P. Amore and F. M. Fern\'andez}{Ritz method for nonlinear problems}

\title{He's amazing calculations with the Ritz method}

\author{Paolo Amore\footnote{paolo.amore@gmail.com}}

\address{ Facultad de Ciencias, Universidad de Colima, Bernal D\'iaz del
Castillo 340, Colima, Colima, Mexico }

\author{Francisco M. Fern\'andez\footnote{fernande@quimica.unlp.edu.ar}}

\address{INIFTA (UNLP, CCT La Plata-CONICET), Divisi\'{o}n
Qu\'{i}mica Te\'{o}rica,
Diag. 113 y 64 (S/N), Sucursal 4, Casilla de Correo 16,
1900 La Plata, Argentina}

\maketitle

\begin{abstract}
We discuss an earlier application of the Ritz variational method for
strongly nonlinear problems. We clearly prove that the results derived for
several extremely simple problems of supposedly physical and mathematical
interest do not provide any clue on the utility of the approach.
\end{abstract}

\section{Introduction}

In a recent review in this journal, He\cite{H06} analyzed several asymptotic
methods for strongly nonlinear equations. One such approach, the so called
Ritz method, consists mainly in converting the nonlinear differential
equation into a Newton equation of motion. Thus, by minimization of the
action integral for the Lagrangian function one obtains an approximate
solution to the nonlinear equation that is expected to be optimal from a
variational point of view. Obviously, the accuracy of the approximate
solution depends heavily on the chosen variational ansatz or trial
``trajectory''.

The purpose of this paper is to analyze the Ritz method proposed by He\cite
{H06} and determine if it is already useful for solving actual nonlinear
problems. For the sake of clarity in what follows we devote a section to
each of the problems discussed.

\section{Anharmonic oscillator}

First, He\cite{H06} transforms the equation of motion for the Duffing
oscillator
\begin{equation}
\ddot{u}(t)-u(t)+\epsilon u(t)^{3}=0  \label{eq:Duffing_eq}
\end{equation}
into the variational integral
\begin{equation}
J(u)=\int \left( -\frac{1}{2}\dot{u}^{2}-\frac{1}{2}u^{2}+\frac{\epsilon }{4}%
u^{4}\right) \,dt
\end{equation}
and concludes that ``it requires that the potential $V(u)=-u^{2}/2+\epsilon
u^{4}/4$ must be positive for all $t>0$, so an oscillation about the origin
will occur only if $\epsilon A^{2}>2$, where $A$ is the amplitude of the
oscillation''. Unfortunately, He\cite{H06} does not show the trial function
from which he draws that conclusion.

The Duffing oscillator has been widely studied and, consequently, its
properties are well known\cite{N81}. For example, from straightforward
inspection of the potential $V(u)$ we already know that there is an unstable
equilibrium point at $u=0$ when $\epsilon <0$. On the other hand, when $%
\epsilon >0$ the potential $V(u)$ exhibits a local maximum $V=0$ at $u=0$
and two minima of depth $-1/(4\epsilon )$ symmetrically located at $\pm 1/%
\sqrt{\epsilon }$. If the initial conditions are such that $V(A)>0$ then the
oscillation will certainly be about the origin; otherwise there will be
oscillations about one of the two minima. It is clear that by simple
inspection of the potential we obtain much more information that the one
derived by He\cite{H06} from the action integral. Therefore, He's
application of the Ritz method to this model is of no relevance whatsoever.

\section{A chemical reaction}

He\cite{H06} proposed the application of the Ritz method to the chemical
reaction
\begin{equation}
nA\rightarrow C+D  \label{eq:ChemReac}
\end{equation}
If $N_{A}(t)$, $N_{B}(t)$, and $N_{C}(t)$ are the number of molecules of the
species $A$, $B$, and $C$, respectively, at time $t$ then He\cite{H06}
assumed that $N_{A}(0)=a$, and $N_{B}(0)=N_{C}(0)=0$. If we call $%
x=N_{B}(t)=N_{C}(t)$, then we conclude that $N_{A}(t)=a-nx$, where $x$ is
known as the extent of reaction\cite{AD02,B60}. The unique rate of reaction
can be defined in terms of the extent of reaction as $v=dx/dt$. He\cite{H06}
further assumed that the rate law is given by
\begin{equation}
\frac{dx}{dt}=k(a-x)^{n}  \label{eq:rate_law}
\end{equation}
At this point we stress the fact that this expression is correct only if the
chemical reaction (\ref{eq:ChemReac}) is elementary, otherwise the rate law
may be more complicated. Most chemical reactions are not elementary and
therefore the reaction order and molecularity do not necessarily agree, as
discussed in any book on physical chemistry\cite{AD02} or chemical kinetics%
\cite{B60}. What is more, the order of reaction may not even be a positive
integer\cite{AD02,B60}. For concreteness here we assume that the rate law (%
\ref{eq:rate_law}) is correct.

He\cite{H06} obtained an approximate solution to the differential equation (%
\ref{eq:rate_law}) by means of the action integral
\begin{equation}
J=\frac{1}{2}\int_{0}^{\infty }\left[ \left( \frac{dx}{dt}\right)
^{2}+k^{2}(a-x)^{2n}\right] dt  \label{eq:J}
\end{equation}
and the variational ansatz
\begin{equation}
x^{var}=a\left( 1-e^{-\eta t}\right)  \label{eq:x_var}
\end{equation}
where $\eta $ is a variational parameter. Notice that $x^{var}(t)$ satisfies
the boundary conditions at $t=0$ and $t\rightarrow \infty $. He\cite{H06}
found that the optimal value of the effective first--order rate constant $%
\eta $ was given by
\begin{equation}
\eta =\frac{ka^{n-1}}{\sqrt{n}}  \label{eq:eta}
\end{equation}
Furthermore, He\cite{H06} argued that chemists and technologists always want
to know the half--time $t_{1/2}=t(x=a/2)$ (which he called halfway time).
According to equations (\ref{eq:x_var}) and (\ref{eq:eta}) the half--time is
given approximately by
\begin{equation}
t_{1/2}^{var}=-\frac{\sqrt{n}\ln (1/2)}{ka^{n-1}}  \label{eq:halftime_var}
\end{equation}

According to He\cite{H06} the exact reaction extent for $n=2$ is
\begin{equation}
x_{He}^{exact}(n=2)=a\left( 1-\frac{1}{1-kat}\right)  \label{eq:x(n=2)_He}
\end{equation}
This result is obviously wrong because it exhibits an unphysical pole at $%
t=1/(ka)$. From this incorrect expression He\cite{H06} derived a meaningless
negative half--time
\begin{equation}
t_{1/2}^{He}(n=2)=-\frac{1}{ka}  \label{eq:halftime_He_exact_n=2}
\end{equation}
In order to obtain a reasonable agreement with the variational result (\ref
{eq:halftime_var}) He\cite{H06} then carried out the following wrong
calculation
\begin{equation}
t_{1/2}^{var}(n=2)=-\frac{\sqrt{2}\ln (1/2)}{ka^{n-1}}=-\frac{0.98}{ka}
\label{eq:halftime_He_var_n=2}
\end{equation}
In this way He\cite{H06} managed to obtain two unphysical negative
half--times that agreed 98\%.

Disregarding the mistakes outlined above we may ask ourselves whether the
approximate variational result may be of any utility to a chemist. Any
textbook on physical chemistry\cite{AD02} or chemical kinetics\cite{B60}
shows that the exact solution to equation (\ref{eq:rate_law}) is
\begin{equation}
x^{exact}=a\left\{ 1-\frac{1}{\left[ 1+k(n-1)a^{n-1}t\right] ^{1/(n-1)}}%
\right\} ,\,n\neq 1  \label{eq:x_exact}
\end{equation}
and that the exact half--time is given by
\begin{equation}
t_{1/2}^{exact}=\frac{2^{n-1}-1}{k(n-1)a^{n-1}}  \label{eq:halftime_exact}
\end{equation}
It is common practice in chemistry to estimate the half--time from
experimental data in order to determine the order of the reaction.
Obviously, an inaccurate expression would lead to an inexact order of
reaction.

The variational half--time (\ref{eq:halftime_var}) is reasonably accurate
for $n=2$ because it is exact for $n=1$. The reason is that the variational
ansatz (\ref{eq:x_var}) is the exact solution for a first--order reaction
when $\eta =k$. Notice that equation (\ref{eq:eta}) leads to such a result
when $n=1$. We can easily verify that the ratio $%
t_{1/2}^{var}/t_{1/2}^{exact}$ increasingly deviates from unity as $n$
increases. Therefore, $n=2$ (the only case selected by He\cite{B60}) is the
most favorable case if $n$ is restricted to positive integers greater than
unity.

The half--time (or half--life) is a particular case of partial reaction
times. We may, for example, calculate the time $t=t_{1/4}$ that has to
elapse for the number of $A$ molecules to reduce to $a/4$ ($x=3a/4$). It is
not difficult to verify that
\begin{equation}
\frac{t_{1/4}^{exact}}{t_{1/2}^{exact}}=2^{n-1}+1
\label{eq:tcuart/tmed_exact}
\end{equation}
From the experimental measure of $t_{1/2}$ and $t_{1/4}$ chemists are able
to obtain the reaction order $n$. However, if they used He's variational
expression (\ref{eq:x_var}) they would obtain
\begin{equation}
\frac{t_{1/4}^{var}}{t_{1/2}^{var}}=2  \label{eq:tcuart/tmed_var}
\end{equation}
that is useless for $n\neq 1$. According to what we have said above it is
not surprising that this ratio is exact for $n=1$. We clearly appreciate
that the variational result does not provide the kind of information that
chemists would like to have because it only predicts first--order reactions.

From the discussion above we conclude that no chemist will resort to the
variational expressions in the study of chemical reactions. There is no
reason whatsoever for the use of an unreliable approximate expression when
one has a simple exact analytical one at hand. Besides, we have clearly
proved that the variational expressions are utterly misleading.

\section{Lambert equation}

He\cite{H06} also applied the Ritz method to the Lambert equation
\begin{equation}
y^{\prime \prime }(x)+\frac{k^{2}}{n}y(x)=(1-n)\frac{y^{\prime }(x)^{2}}{y(x)%
}  \label{eq:Lambert_y}
\end{equation}
and arrived at the variational formulation
\begin{equation}
J(y)=\frac{1}{2}\int \left( -n^{2}y^{2n-2}y^{\prime 2}+k^{2}y^{2n}\right)
\,dt
\end{equation}
By means of the transformation $z=y^{n}$ He obtained
\begin{equation}
J(z)=\frac{1}{2}\int \left( -z^{\prime 2}+k^{2}z^{2}\right) \,dt
\end{equation}
that leads to the Euler--Lagrange equation
\begin{equation}
z^{\prime \prime }+k^{2}z=0  \label{eq:Lambert_z}
\end{equation}
Obviously, the solution to this linear equation is straightforward.

If we substitute the transformation $z=y^{n}$ into equation (\ref
{eq:Lambert_y}) we obtain equation (\ref{eq:Lambert_z}) in a more direct
way. Therefore, there is no necessity for the variational Ritz method.

\section{Soliton solution}

He\cite{H06} also studied the KdV equation
\begin{equation}
\frac{\partial u(x,t)}{\partial t}-6u(x,t)\frac{\partial u(x,t)}{\partial x}+%
\frac{\partial ^{3}u(x,t)}{\partial x^{3}}=0  \label{eq:Kdv}
\end{equation}
and looked for its travelling--wave solutions in the frame
\begin{equation}
u(x,t)=u(\xi ),\,\xi =x-ct  \label{eq:KdV_frame}
\end{equation}
The function $u(\xi )$ satisfies the nonlinear ordinary differential
equation
\begin{equation}
u^{\prime \prime \prime }(\xi )-cu^{\prime }(\xi )-6u(\xi )u^{\prime }(\xi
)=0  \label{eq:KdV_2}
\end{equation}
where the prime indicates differentiation with respect to $\xi $. Then He%
\cite{H06} integrated this equation (taking the integration constant
arbitrarily equal to zero) and obtained
\begin{equation}
u^{\prime \prime }(\xi )-cu(\xi )-3u(\xi )^{2}=0  \label{eq:KdV_3}
\end{equation}
By means of the so called semi--inverse method He\cite{H06} obtained the
variational integral
\begin{equation}
J=\int_{0}^{\infty }\left[ \frac{1}{2}\left( \frac{du}{d\xi }\right) ^{2}+%
\frac{c}{2}u^{2}+u^{3}\right] \,dt
\end{equation}
Choosing the trial function
\begin{equation}
u=p\cosh ^{-2}(q\xi )  \label{eq:KdV_trial}
\end{equation}
where $p$ and $q$ are variational parameters, He\cite{H06} obtained $p=c/2$
and $q=\sqrt{c}/2$.

By substitution of equation (\ref{eq:KdV_trial}) into equation (\ref
{eq:KdV_2}) we obtain the same values of $p$ and $q$ in a more direct way
and with less effort. Therefore, there is no need for the variational method
for the successful treatment of this problem.

\section{Bifurcation}

He\cite{H06} also applied the Ritz method to the most popular Bratu equation
\begin{equation}
u^{\prime \prime }(x)+\lambda e^{u(x)}=0,\,u(0)=u(1)=0  \label{eq:Bratu}
\end{equation}
that has been studied by several authors\cite{W89} (and references therein).
Here we only cite those papers that are relevant to present discussion. He%
\cite{H06} derived the action integral
\begin{equation}
J=\int_{0}^{1}\left( \frac{1}{2}u^{\prime 2}-\lambda e^{u}\right) \,dx
\label{eq:Bratu_int}
\end{equation}
and proposed the simplest trial function that satisfies the boundary
conditions:
\begin{equation}
u(x)=Ax(1-x)  \label{eq:Bratu_trial}
\end{equation}

Curiously, He\cite{H06} appeared to be unable to obtain an exact analytical
solution for the integral; however, it is not difficult to show that
\begin{equation}
J(A)=\frac{A^{2}}{6}-\sqrt{\frac{\pi }{A}}\lambda e^{A/4}\mathop{\rm erf}%
\left( \sqrt{\pi }/2\right)  \label{eq:Bratu_J(A)}
\end{equation}
We cannot exactly solve $dJ(A)/dA=0$ for $A$ but we can solve it for $%
\lambda $:
\begin{equation}
\lambda =\frac{4A^{5/2}}{3\left[ \sqrt{\pi }(A-2)e^{A/4}\mathop{\rm erf}%
\left( \sqrt{A}/2\right) +2\sqrt{A}\right] }  \label{eq:Bratu_lambda}
\end{equation}
The analysis of this expression shows that $\lambda (A)$ exhibits a maximum $%
\lambda _{c}=3.569086042$ at $A_{c}=4.727715383$. Therefore there are two
variational solutions for each $0<\lambda <\lambda _{c}$, only one for $%
\lambda =\lambda _{c}$ and none for $\lambda >\lambda _{c}$. This conclusion
agrees with the rigorous mathematical analysis of the exact solution\cite
{W89} that we will discuss below. Besides, the critical value of the
adjustable parameter $A_{c}$ is also a root of $d^{2}J(A)/dA^{2}=0$.

The exact solution to the one--dimensional Bratu equation~(\ref{eq:Bratu})
is well--known. Curiously enough, He\cite{H06}, Deeba et al\cite{DKX00}, and
Khury\cite{K04} showed a wrong expression. A correct one is (notice that one
can write it in different ways)
\begin{equation}
u(x)=-2\ln \left\{ \frac{\cosh \left[ \theta (x-1/2)\right] }{\cosh (\theta
/2)}\right\}  \label{eq:Bratu_exact}
\end{equation}
where $\theta $ is a solution to
\begin{equation}
\lambda =\frac{2\theta ^{2}}{\cosh (\theta /2)^{2}}  \label{eq:lambda(theta)}
\end{equation}
The critical $\lambda $--value is the maximum of $\lambda (\theta )$, and we
easily obtain it from the root of $d\lambda (\theta )/d\theta =0$ that is
given by
\begin{equation}
e^{\theta _{c}}(\theta _{c}-2)-\theta _{c}-2=0  \label{eq:theta_c}
\end{equation}
The exact critical parameters are $\theta _{c}=2.399357280$ and $\lambda
_{c}=3.513830719$ that lead to $u^{\prime }(0)_{c}=4$. We appreciate that
the variational approach provides a reasonable qualitative (or even semi
quantitative) description of the problem.

We may try a perturbation approach to the Bratu equation in the form of a
Taylor series in the parameter $\lambda $:
\begin{equation}
u(x)=\sum_{j=0}^{\infty }u_{j}(x)\lambda ^{j}  \label{eq:PT_u(x)}
\end{equation}
where, obviously, $u_{0}(x)=0$. From the exact expression we obtain
\begin{equation}
u^{\prime }(0)=\frac{\lambda }{2}+\frac{\lambda ^{2}}{24}+\frac{\lambda ^{3}%
}{160}+\ldots \approx 0.5\lambda +0.0417\lambda ^{2}+0.00625\lambda
^{3}+\ldots  \label{eq:PT_u'_exact}
\end{equation}
while the variational approach also yields a reasonable result
\begin{equation}
u^{\prime }(0)=\frac{\lambda }{2}+\frac{\lambda ^{2}}{20}+\frac{43\lambda
^{3}}{5600}+\ldots \approx 0.5\lambda +0.05\lambda ^{2}+0.00768\lambda
^{3}+\ldots  \label{eq:PT_u'_He}
\end{equation}
It seems that the Ritz method already produces satisfactory results for this
kind of two--point boundary value problems.

Another simple variational function that satisfies the same boundary
conditions is
\begin{equation}
u(x)=A\sin (\pi x)  \label{eq:Bratu_trial_sin}
\end{equation}
It leads to the following variational integral:
\begin{equation}
J(A)=\frac{A^{2}\pi ^{2}}{4}-\lambda \left[ I_{0}(A)+L_{0}(A)\right]
\label{eq:Bratu_J(A)_2}
\end{equation}
where $I_{\nu }(z)$ and $L_{\nu }(z)$ stand for the modified Bessel and
Struve functions\cite{AS72}, respectively. From the minimum condition we
obtain
\begin{equation}
\lambda =\frac{A\pi ^{3}}{2\left\{ 2+\pi \left[ I_{1}(A)+L_{1}(A)\right]
\right\} }  \label{eq:Bratu_lambda_2}
\end{equation}
It is not difficult to show that this trial function yields better critical
parameters: $\lambda _{c}=3.509329130$ and $u^{\prime }(0)_{c}=3.756549365$.
Besides, one can easily derive the approximate perturbation expansion
exactly
\begin{eqnarray}
u^{\prime }(0) &=&\frac{4\lambda }{\pi ^{2}}+\frac{4\lambda ^{2}}{\pi ^{4}}+%
\frac{4\left( 3\pi ^{2}+16\right) \lambda ^{3}}{3\pi ^{8}}+\frac{4\left( \pi
^{2}+18\right) \lambda ^{4}}{\pi ^{10}}+\ldots  \nonumber \\
&\approx &0.405\lambda +0.0411\lambda ^{2}+0.00641\lambda ^{3}+\ldots
\label{eq:PT_u'_present}
\end{eqnarray}
Notice that although the coefficient of $\lambda $ is not exact the
remaining ones are more accurate than those of the preceding trial function.

Fig.~\ref{fig:Bratu} shows the exact slope at origin $u^{\prime }(0)$ in
terms of $\lambda $ and the corresponding estimates given by the two
variational functions. We appreciate that the variational approach proposed
by He\cite{H06} yields the solution with smaller $u^{\prime }(0)$ (lower
branch) more accurately than the other one (upper branch). This comparison
between the exact and approximate solutions for a wide range of values of $%
\lambda $ was not carried out before; He\cite{H06} simply compared the two
slopes at origin for just $\lambda =1$. The other trial function~(\ref
{eq:Bratu_trial_sin}) yields a better overall approximation at the expense
of the accuracy for small values of $\lambda $. Some more elaborated
approaches, like the Adomian decomposition method, fail to provide the upper
branch\cite{DKX00}; therefore, the Ritz method seems to be suitable for the
analysis of this kind of nonlinear problems.

We may conclude that the Ritz variational method provides a useful insight
into the Bratu equation. However, one should not forget that there is a
relatively simple exact solution to this problem and that the generalization
of the approach in the form of a power series proposed by He\cite{H06}
\begin{equation}
u(x)=Ax(1-x)(1+c_{1}x+c_{2}x^{2}+\ldots )  \label{eq:Bratu_trial_power}
\end{equation}
may surely lead to rather analytically intractable equations.

\section{Conclusions}

Historically, scientists have developed perturbational, variational and
numerical approaches to solve nontrivial mathematical problems in applied
mathematics and theoretical physics. In some cases, where the exact solution
exists but is given by complicated special functions, an approximate simpler
analytical solution may nonetheless be of practical utility. However, He\cite
{H06} chose examples where either the Ritz method does not provide any
useful insight, or the exact analytical solutions are as simple as the
approximate ones, or the direct derivation of the exact result is more
straightforward than the use of the variational method. From the discussions
in the preceding sections we may conclude that He's application of the Ritz
variational method\cite{B60} does not show that the approach is suitable for
the treatment of nonlinear problems. In most of the cases studied here the
straightforward analysis of the problem yields either more information or
the same result in a more direct way.

\begin{figure}[H]
\begin{center}
\includegraphics[width=9cm]{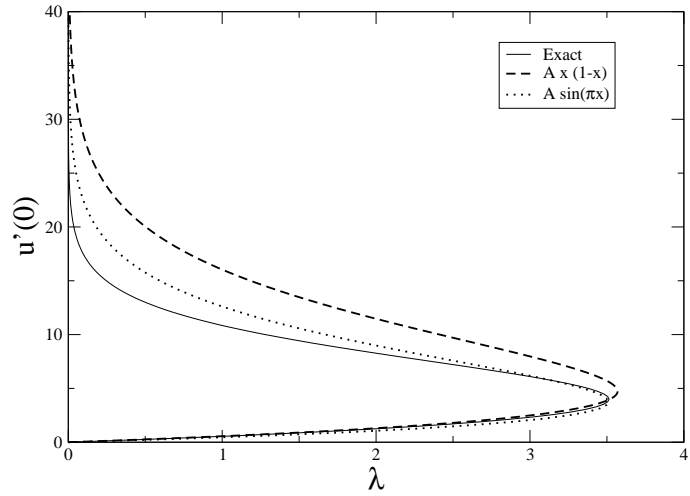}
\end{center}
\caption{Bifurcation diagram for the slope at origin $u^\prime (0)$ in terms
of $\lambda$}
\label{fig:Bratu}
\end{figure}

To be fair we should mention that the Ritz variational method provides a
reasonable bifurcation diagram by means of relatively simple trial functions
as shown in Fig.~\ref{fig:Bratu}. However, even in this case the utility of
the approach is doubtful because there exists a remarkably simple analytical
solution to that equation. The treatment of a nontrivial example is
necessary to assert the validity of the approach.

He's choice of the rate equation for chemical reactions\cite{H06} is by no
means a happy one (without mentioning the mistakes in the calculations). In
this case the exact solution is quite simple and the variational ansatz is
unsuitable for practical applications. We may argue that a trial function
with the correct asymptotic behaviour would yield meaningful results. In
fact, it may even produce the exact result; but one should not forget that
such a success would obviously be due to the fact that there exist a
remarkably simple exact solution available by straightforward integration.

As said before, present results show that He\cite{H06} failed to prove that
the Ritz variational method provides a successful way of treating strongly
nonlinear problems. Of course, the main ideas behind that variational method
are correct, and the case of Bratu equation suggests that it may be possible
to find appropriate trial functions for the successful treatment of some
problems. Unfortunately, the remaining He's choices\cite{H06} do not do much
to convince one that the method is worthwhile.

He's article\cite{H06} is an example of the kind of poor research papers
that have been lately published by some supposedly respectable journals. It
is part of such journals' policy to reject comments that can reveal this
unfortunate situation. One may ask oneself what is the profit that those
journals get from such a practice. If the reader is interested in other
examples of poor research papers in supposedly respectable journals I
suggest some earlier reports in this
forum\cite{F07,F08f,F08e,F08d,F08c,F08b}.

\end{document}